\documentstyle[epsfig,twocolumn,pra,aps,floats]{revtex}
\draft

\begin{document}
\title{Tuning the dipolar interaction in quantum gases}
\author{Stefano Giovanazzi, Axel G\"{o}rlitz and Tilman Pfau}
\address{5th Institute of Physics, University of Stuttgart, D-70550 Stuttgart, Germany.}
\date{\today}
\maketitle
\begin{abstract}
We have studied the tunability of the interaction between
permanent dipoles in Bose-Einstein condensates. Based on
time-dependent control of the anisotropy of the dipolar
interaction, we show that even the very weak magnetic dipole
coupling in alkali gases can be used to excite collective modes.
Furthermore, we discuss how the effective dipolar coupling in a
Bose-Einstein condensate can be tuned from positive to negative
values and even switched off completely by fast rotation of the
orientation of the dipoles.
\end{abstract}

In the Bose-Einstein condensates (BECs) created thus far
\cite{stringari} atoms interact essentially only at very short
distance. At low temperatures, the interaction between these
neutral atoms can be typically characterized by a single
parameter, the s-wave scattering length. The magnitude and sign of
the s-wave scattering length can be changed by tuning the external
magnetic fields in the vicinity of a so-called 'Feshbach
resonance' \cite{inouye98}. This technique has opened new
possibilities for the study and manipulation of BECs. In
particular, the value of the s-wave scattering length can be
varied in real time, allowing, for example, the observation of
collapsing and exploding BECs \cite{donley01}.

In  recent years an increasing interest is devoted to
dipole-dipole interactions in ultracold gases
\cite{yi00,goral00,santos00,baranov,goral,meystre01,deMille,giova02,santos02}.
Dipolar interactions would largely enrich the variety of phenomena
to be observed due to their long-range and anisotropic  character.
If atomic dipole moments are sufficiently large, the resulting
dipole-dipole forces may influence, or even completely change the
properties of Bose gases \cite{yi00,goral00,santos00,giova02}, the
conditions for BCS transition in Fermi gases \cite{baranov}, or
the phase diagram for quantum phase transitions in ultracold
dipolar gases confined in optical lattices \cite{goral}. The
interplay of short-range scattering and long-range interaction may
give rise to phenomena like ferromagnetic order and spin waves
\cite{meystre01}. Moreover the dipolar particles are considered to
be promising candidates for the implementation of fast and robust
quantum-computing schemes \cite{deMille}.

In this Letter, we consider the case of permanent (magnetic)
dipoles. For alkali atoms with a magnetic moment $m$ of $\mu_B$,
where $\mu_B$ is a Bohr magneton, the long range part of the
magnetic dipole-dipole interaction is generally neglected since it
is very small compared to the s-wave pseudo-potential. Here, we
show that even in alkali BECs this dipolar interaction can be made
visible if the dipole-dipole interaction is modulated by rotating
the atomic dipoles.  For appropriate rotation frequencies, a
coupling to elementary excitations can be achieved leading to a
dynamic growth of the amplitude of the excitation. In particular,
we will show how the dipolar coupling in an alkali condensate can
be used to excite the quadrupole mode and how it is possible to
extract quantitative information about the ratio of the dipolar
coupling and the usual s-wave coupling.

In ensembles of atoms with a larger magnetic moment, such as
chromium where $m = 6 \mu_B$, the magnetic dipole-dipole
interaction energy may become comparable in size with the s-wave
contribution to the mean-field energy. Therefore, even the static
properties of a condensate such as the aspect ratio are
significantly altered. Here, we show how the effective strength of
the interaction and thus its effect on the static properties of a
condensate can be varied by fast rotation of the atomic dipoles.
In the case of a dipolar interaction which is stronger than the
s-wave interaction only certain combinations of trap geometry and
orientation of the dipoles lead to a stable (or metastable)
condensate \cite{santos00,yi00}. By tuning the dipolar interaction
unstable condensates can thus be transformed into stable ones and
viceversa.  In this paper, we study the interaction between
permanent magnetic dipoles. However, the methods we propose can
also be adapted for particles with electric dipole moment such as
heteronuclear  molecules. In this case the dipolar interaction is
expected to dominate over the s-wave interaction and thus the
tuning of the interaction might be necessary to reach stability.

\emph {Manipulating the magnetic dipolar coupling in a BEC}---The
long-range part of the interaction between two magnetic dipole
moments ${\bf m}_1$ and ${\bf m}_2$ of atoms located at ${\bf
r}_1$ and ${\bf r}_2$ takes the familiar dependence
\begin{eqnarray}
U_{\mathrm{dd}}({\bf r})= - \frac{\mu_0}{4 \pi} \, {3\, ({\bf m}_1
\cdot \hat{{\bf r}}) \, ({\bf m}_2 \cdot \hat{{\bf r}}) - ({\bf
m}_1 \cdot {\bf m}_2) \over r^3} \label{interaction0}
\end{eqnarray}
as a function of the interatomic distance  ${\bf r} = {\bf r}_2 -
{\bf r}_1$, where $\hat{{\bf r}} = {\bf r}/r$ and $\mu_0$ is the
magnetic permeability of the vacuum. We consider the case of a
general \emph{time-dependent} homogeneous magnetic field
\begin{eqnarray}
{\bf B}(t)&=& B\,{\bf e}(t)\;,\label{bfieldgen}
\end{eqnarray}
slowly-varying with respect to the Larmor frequency
$\omega_{Larmor}=m B/\hbar$, where $m=|{\bf m}_1|=|{\bf m}_2|$ for
identical atoms. A spin-polarized atomic ensembles will
adiabatically follow the external magnetic field $B(t)$ and the
resulting interatomic energy becomes
\begin{eqnarray}
U_{\mathrm{dd}}({\bf r},t)= - \frac{\mu_0 m^2}{4 \pi} \, {3\,
({\bf e} (t) \cdot \hat{{\bf r}})^2 - 1 \over r^3}
\label{interactionrotating}
\end{eqnarray}
and is thus \emph{time-dependent}.

Our analysis is based on the mean-field approach in the
Thomas-Fermi limit already used in the context of dipolar Bose
gases
\cite{yi00,goral00,santos00,baranov,goral,meystre01,deMille,giova02,santos02}.
Such a description can be accomplished through the hydrodynamic
equations for superfluids \cite{stringari}:
\begin{eqnarray}
{\partial n \over \partial t} &=&- {\bf \nabla}(n {\bf v})  \;,\label{hydro1}\\
{\partial {\bf v} \over \partial t} &=&- {\bf \nabla} \left( \frac
{v^2}{2} + {\delta\mu\over M} \right),\label{hydro2}
\end{eqnarray}
where $n$ is the density, ${\bf v}$ the velocity field and
$\delta\mu$  the difference between the mean-field plus external
fields and the chemical potential $\mu$ given by
\begin{eqnarray}
\delta\mu &=&  g n  + \Phi_{\mathrm{dd}} + {M \omega^2_{0}
r^2\over 2} - \mu .\label{hydro3}
\end{eqnarray}
Here $M$ is the atomic mass,  $M \omega_{0}^2 r^2/ 2$ is an
isotropic  harmonic potential which could be created using optical
dipole forces, $g=4\pi\hbar^2a/M$ and $a$ is the s-wave scattering
length which we assume to be positive. $\Phi_{\mathrm{dd}}({\bf
r},t)$ is the mean-field potential generated by the (time
dependent) dipolar interaction
\begin{eqnarray}
\Phi_{\mathrm{dd}}({\bf r},t)= \int d{\bf r}' \,
U_{\mathrm{dd}}({\bf r}-{\bf r}',t) \,n({\bf r}') \;.
\end{eqnarray}
In the Thomas-Fermi limit it is useful to define a dimensionless
quantity
\begin{eqnarray}
\varepsilon_{\mathrm{dd}} = \frac{\mu_0 m^2 M}{12 \pi \hbar^2
a}\;,\label{epsilon}
\end{eqnarray}
which is a measure of the strength of the dipole-dipole
interaction relative to the s-wave scattering energy.
Modifications in the ground state condensate density can only be
expected if $\varepsilon_{\mathrm{dd}}$ is appreciably different
from zero.

For chromium the s-wave scattering length $a_{\mathrm{Cr}}$ is not
known. If we assume that $a_{\mathrm{Cr}}$ is equal to the sodium
scattering length ($a=2.8$ nm) the dipolar strength parameter
becomes $\varepsilon_{\mathrm{dd}}^{\mathrm{Cr}}=0.29$
($m=6\mu_B$). Therefore, the dipolar force is expected to lead to
visible modification in the density of a chromium BEC.
Differently, in both rubidium and sodium the magnetic dipolar
energy is rather small compared to the mean-field s-wave
scattering energy. The values of $\varepsilon_{\mathrm{dd}}$ are
$\varepsilon_{\mathrm{dd}}^{\mathrm{Rb}^{87}}=0.0064$ and
$\varepsilon_{\mathrm{dd}}^{\mathrm{Na}}=0.0035$, in the doubly
spin-polarized ground-state, where $m=\mu_B$. For heteronuclear
molecules with a permanent electric dipole moment of one Debye the
corresponding dipolar coupling will be increased by
approximatively the inverse square of the fine structure constant.
This implies that $\varepsilon_{\mathrm{dd}}$ \cite{expression} is
on the order of $10^2$.

\emph {Probing the quadrupole modes in sodium and rubidium}---The
time modulation of the dipolar coupling
(\ref{interactionrotating}) can be used to resonantly excite a
quadrupole mode even for sodium and rubidium condensates. For
simplicity we restrict ourselves to the case of a condensate in
the Thomas-Fermi regime confined in an isotropic  harmonic trap
with frequency $\omega_0$ (for example $\omega_0 / 2 \pi =  100$
Hz). Because $\varepsilon_{\mathrm{dd}}$ is small the
dipole-dipole mean-field potential $\Phi_{\mathrm{dd}}$ can be
calculated using the symmetric equilibrium density distribution
$n_0 = M\omega_0^2 (R^2-r^2)/2g$, where $R$ is its radius
\begin{eqnarray}
\Phi_{\mathrm{dd}}({\bf r},t)&=&-
\frac{\varepsilon_{\mathrm{dd}} M \omega_0^2}{5} \, \left[3\,
({\bf e} (t) \cdot \hat{{\bf r}})^2 - 1\right] r^2 \;\;\; r<R,
\label{linear1}\\
&=&- \frac{\varepsilon_{\mathrm{dd}} M \omega_0^2}{5} \, \left[3\,
({\bf e} (t) \cdot \hat{{\bf r}})^2 - 1\right]{R^5\over
r^3}\;\;\;r>R. \label{linear2}
\end{eqnarray}
The potential outside the condensate region ($r>R$) corresponds
exactly to the field generated by N dipoles located in the center
of the condensate. Differently, inside the condensate ($r<R$) the
mean-field potential has an anisotropic but harmonic dependence.

If the interaction is driven on a resonance frequency the
evolution of the condensate can be modified even for very small
dipolar coupling. Consider a magnetic field rotating in the plane
$x-y$ as ${\bf B}(t)= B{\bf e}(t)$, where
\begin{eqnarray}
{\bf e}(t)= \cos(\Omega t)\hat{{\bf x}}+\sin(\Omega t)\hat{{\bf
y}} \,.\label{bfield}
\end{eqnarray}
The resulting dipolar interatomic energy
(\ref{interactionrotating}) and its corresponding contribution to
the mean-field potential (\ref{linear1}-\ref{linear2}), will have
a 'rotating' part \cite{armoniche}. For small dipolar coupling
within alkali gases the non oscillating part of (\ref{linear1})
can be neglected and the time-dependent part will resonantly
couple to the 'transversal' quadrupole mode when
$\Omega\approx\omega_0/\sqrt{2}$ (see Ref. \cite{stringari96}).
After linearization of the equations of motion (\ref{hydro1}) and
(\ref{hydro2}) we obtain
\begin{eqnarray}
{\partial^2 \delta n \over \partial t^2} = {\bf \nabla} \left(
\frac{g n_0}{M} {\bf \nabla} \left(\delta n \right) \right) + {\bf
\nabla} \left( \frac{ n_0}{M} {\bf \nabla}
\left({\Phi_{\mathrm{dd}}} \right) \right),\label{hydrolinear}
\end{eqnarray}
where $\delta n=n-n_0$ is the density displacement around the
equilibrium density $n_0$. The time-dependent part of
$\Phi_{\mathrm{dd}}$ corresponds in Eq. (\ref{hydrolinear}) to the
rotation of an anisotropic harmonic trap with deformation
\begin{eqnarray}
\varepsilon = {\omega_X^2 - \omega_Y^2 \over \omega_X^2 +
\omega_Y^2} = {3\over 10} \varepsilon_{\mathrm{dd}}\,,
\label{epsilonsandro}
\end{eqnarray}
where $\omega_X$ and $\omega_Y$ are the maximum and minimum values
of the rotating trap frequencies. Neglecting the time-independent
part of (\ref{linear1}) the density will evolve as
\begin{eqnarray}
\delta n  = - \frac{3\,\varepsilon_{\mathrm{dd}} M
\omega_0^2}{10\, g} \, r^2 \sin^2\theta \sin(2\phi-2\Omega t
)\,\Omega t, \label{soluz}
\end{eqnarray}
where $\phi$ is the azimuthal angle ($r<R$). It is noteworthy that
the rotating part \cite{armoniche} of (\ref{linear1}) couples
\emph{only} the first quadrupole mode in the Thomas-Fermi limit
\cite{stringari96}. From the study of the evolution of the
condensate deformation
\begin{eqnarray}
\alpha = {\langle x^2 - y^2 \rangle \over \langle x^2 + y^2
\rangle} = -{3\over 5} \varepsilon_{\mathrm{dd}}\, \Omega
t\,\sin(2\Omega t) \label{epsilonsandro}
\end{eqnarray}
we can extract information about the ratio of the dipolar coupling
versus the standard mean-field energy. If the quality factor
$Q=\sqrt{2}\omega_0\tau$ of such a resonance, where $\tau$ is the
corresponding decay time, is assumed to be 50 \cite{commento} the
amplitude of the oscillation of the condensate anisotropy $\alpha$
is expected to be $|\alpha|=3 Q \varepsilon_{\mathrm{dd}} /10 \sim
9.5 \times 10^{-2}$ for rubidium (see Fig. 1). We have performed
our calculation in the linear regime and for an isotropic trap.
The non-linearity of the mean-field is expected to give a
significant correction for larger $Q$. The extension of the
calculation to an axially symmetric cigar-shaped trap is expected
to yield similar results.
\begin{figure}[t]
\centerline{\psfig{figure=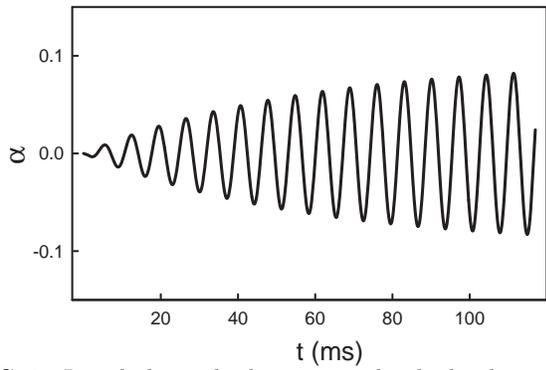,height=1.9in}} \caption{In
rubidium the long-range dipole-dipole interaction does not modify
in an appreciable way the ground state density. When the magnetic
dipoles are driven on resonance with a transversal quadrupole mode
the anisotropy $\alpha$ (Eq. (15)) is growing with time (here we
assume a quality factor $Q=50$).}
\end{figure}

\emph {Tunability of the magnetic dipolar interaction:} --- We
consider a condensate in an axially symmetric harmonic trap, with
radial and axial trap frequencies $\omega_r$ and $\omega_z$. The
magnetic field
\begin{eqnarray}
{\bf B}(t)= B\,\left[\cos(\varphi) \hat{\bf{z}} + \sin(\varphi)
\left(\cos(\Omega t)\hat{\bf{x}}+\sin(\Omega
t)\hat{\bf{y}}\right)\right]\label{bfield}
\end{eqnarray}
is a combination of a static magnetic field $B_z$ directed along
the $z$ direction and a fast rotating field $B_{\rho}$ in the
radial plane. The frequency is chosen such that the atoms are not
significantly moving during the time $\Omega^{-1}$; while the
magnetic moments will follow adiabatically the external field
$\bf{B}(t)$; this corresponds to $\omega_{Larmor}\gg \Omega \gg
\omega_{trap}$, where $\omega_{trap}$ is any of the external trap
frequencies. In this limit we can consider the average of the
interaction (\ref{interactionrotating}) in the period
$2\,\pi/\Omega$, resulting in the cylindrically symmetric
interatomic potential
\begin{eqnarray}
\langle U_{\mathrm{dd}}(\vec r)\rangle &=& - \frac{\mu_0 m^2}{4
\pi} \left({3\,\cos^2\varphi -1\over 2}\right)
\left({3\,\cos^2\theta -1}\right). \label{interactionaveraged}
\end{eqnarray}
The averaged interaction energy (\ref{interactionaveraged}) equals
$(3\cos^2\varphi-1)/2 $ times the interatomic energy of dipoles
aligned along the $z$-axes. Note that this factor can be changed
continuously from $1$ to $-1/2$, providing the possibility to
change the dipolar interaction from attractive to repulsive. At a
particular angle $\varphi_{M}=54.7^o$ the dipolar interaction
averages  to zero. This angle, the so-called magic-angle, is well
known in solid-state nuclear magnetic resonance (NMR) technology
\cite{AndrewLowe59}. The tunability of the dipolar interaction
will allow to observe changes in the condensate eccentricity
which should be a $10\%$ effect in Cr assuming
$a_{\mathrm{Cr}}=a_{\mathrm{Na}}$ or in the collective excitation
frequencies \cite{yi00,goral2}.

\begin{figure}[t]
\centerline{\psfig{figure=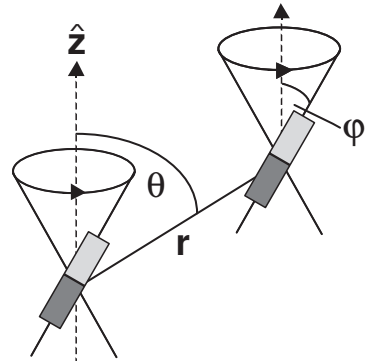,height=1.9in}}
\caption{Tunability of the magnetic dipole interaction. The atomic
dipoles follow adiabatically the total magnetic field
(\ref{bfield}) when the field is sufficiently strong. The angle
$\varphi$ between the precessing total magnetic field and the
$z$-axis characterizes the interaction. $\varphi=0$: the magnetic
dipoles are simply polarized in the $z$ direction.
$\varphi=\pi/2$: the magnetic dipoles follow the rotating magnetic
field and as a result of the fast rotation the averaged
interaction becomes $-1/2$ times the interaction in the polarized
case. $\varphi=54.7^o$ (magic-angle): the dipolar interaction
averages to zero.}
\end{figure}

\emph {Stabilizing strongly interacting dipolar gases:}---Let us
now apply this tunability to the stability diagram of dipolar
BECs. We consider the case in which the dipolar coupling is larger
than the s-wave pseudopotential ($\varepsilon_{\mathrm{dd}}\gtrsim
1$). Here the tunability can stabilize strongly interacting
dipolar gases like heteronuclear molecules which otherwise exhibit
instability over a wide range of parameters. This regime,
considered also in Ref. \cite{yi00,santos00}, may be achieved in a
cloud of magnetic dipolar atoms, by making use of a Feshbach
resonance to reduce the s-wave scattering length.

In a spin polarized cloud the stability diagram as well as the
condensate anisotropy can be studied in the Thomas-Fermi limit as
function of $\epsilon_{\mathrm{dd}}$ (see Ref.
\cite{yi00,goral00,santos00}). When $\epsilon_{\mathrm{dd}}\geq 1$
the cloud may be unstable against collapse, i.e. can be always
possible to find a density configuration of arbitrary large
negative energy. The stability as $\epsilon_{\mathrm{dd}}$ and the
trap anisotropy $l=\omega_z/\omega_r$ are varied can be
investigated in analogy to Ref. \cite{yi00}.

Within the Thomas-Fermi approximation, the stability analysis can
be made using a scaling variational approach \cite{zoller}.
Assuming a density for the condensate of the form
\begin{eqnarray}
n(x,y,z)={1\over \lambda_r^2 \lambda_z}
n_0\left({x\over\lambda_r},{y\over\lambda_r},{z\over\lambda_z}\right),\label{gausan}
\end{eqnarray}
where $\lambda_r$ and $\lambda_z$ are scaling variational
parameters and $n_0$ is a spherically symmetric density
distribution (for instance the usual Thomas-Fermi inverted
parabola). The condensate anisotropy $\lambda_r/\lambda_z$ can be
found by minimizing the expectation value of the total energy $
H_{\mathrm{tot}}= H_{\mathrm{ho}} + H_{\mathrm{int}}$ where
\begin{eqnarray}
H_{\mathrm{ho}}=c_1 N m\left(2 \omega_r^2 \lambda_r^2 +\omega_z^2
\lambda_z^2\right)
\end{eqnarray}
is the harmonic potential energy and
\begin{eqnarray}
H_{int}&=& c_2 {N^2 g \over \lambda_r^2
\lambda_z}\left[1-\varepsilon_{\mathrm{dd}}
\left({3\,\cos^2\varphi - 1\over 2}\right)
f\left({\lambda_r\over\lambda_z}\right)\right] \label{hintvar}
\end{eqnarray}
 the mean-field interaction energy, where $c_1$ and $c_2$ are numerical
constants \cite{c1c2} and
\begin{eqnarray}
f(\kappa)&=&\frac{1+2\kappa^2}{1-\kappa^2}-\frac{3\kappa^2
{\tanh}^{-1}\sqrt{1-\kappa^2}}{\left(1-\kappa^2\right)^{3/2}}
\end{eqnarray}
derives from an angular integration.

The stability can be varied for a fixed value of
$\varepsilon_{\mathrm{dd}}$ and the trap anisotropy $l$ just
varying one easy experimental parameter, namely the angle
$\varphi$. In Fig. 3 we show one example of how the stability
phase diagram for $\varepsilon_{\mathrm{dd}}=4$ can be varied as
function of the angle $\varphi$. For very large
$\varepsilon_{\mathrm{dd}}$ the stable region in the $\varphi$-$l$
plane becomes a thin region around the magic angle line.

\begin{figure}[t]
\centerline{\psfig{figure=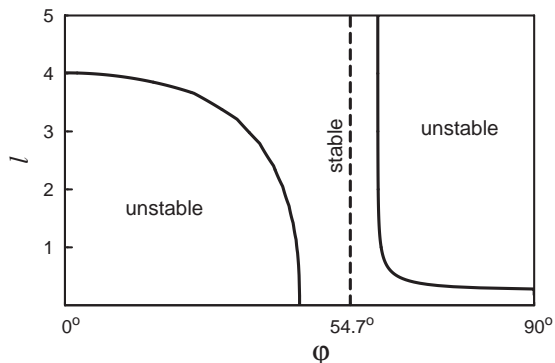,height=1.9in}}
\caption{Stability phase diagram in the $\varphi$-$l$ plane
($l=\omega_z/\omega_r$) obtained in the Thomas-Fermi limit
($N\rightarrow\infty$) and with $\varepsilon_{\mathrm{dd}}=4$.
Condensate are always stable in the vicinity of the magic angle
$\varphi_{M}=54.7^o$.}
\end{figure}

\emph { Conclusion}---We have shown that the anisotropy of the
dipole-dipole interaction of a polarized gas can be used to tune
the strength of the dipolar coupling and to excite collective
excitations. 'Tuning' the dipolar coupling will become an
important tool for 'designing' atomic quantum gases with novel
properties and dynamical aspects. This approach in optically
trapped Bose condensed gases should allow to make even very weak
dipolar coupling visible with experimentally feasible parameters
(like $B$ fields in the Gauss region and spinning frequencies up
to several $10$ kHz). The tuning of the interaction might become a
relevant technique for stabilizing strongly interacting dipolar
systems like heteronuclear molecular gases.

\emph {Acknowledgement:}---Valuable discussion with G. Denninger
are acknowledged; funding was provided by the RTN network "cold
quantum gases" under the contract number HPRN-CT-2000-00125, and
the Deutsche Forschungsgemeinschaft.


\begin{references}
\bibitem{expression} One can define an expression analogous to Eq.
(\ref{epsilon}) for 'permanent' electric dipoles as
$\varepsilon_{\mathrm{dd}} = \frac{ d^2 M}{12 \pi \epsilon_0
\hbar^2 a}$ where $d$ is the electric permanent dipole moment and
$a$ a scattering length.
\bibitem{c1c2} $c_1$ and $c_2$ depend only on $n_0$;
because $c_1$ and $c_2$ can be reabsorbed in a global re-scaling
of the energy and variational parameters, the study of the
condensate anisotropy and stability does not depend on $n_0$.
\end{references}
\end{document}